\documentclass{ws-ijmpa}
\usepackage[super,compress]{cite}
\usepackage{graphicx}
\newcommand{\p}[1]{(\ref{#1})}
\begin{document}
\markboth{A.A. Isayev} {Absolute stability window and upper bound on
the magnetic field\dots}

%
\catchline{}{}{}{}{}
%

\title{ABSOLUTE STABILITY WINDOW AND UPPER BOUND  ON THE MAGNETIC FIELD
STRENGTH IN A STRONGLY MAGNETIZED STRANGE QUARK STAR  }

\author{A. A. ISAYEV
}
\address{
Kharkov Institute of Physics and Technology,  \\ Academicheskaya
Street 1,
 Kharkov, 61108, Ukraine \\
Kharkov National University, Svobody Sq., 4, Kharkov, 61022, Ukraine \\
Institute for the Early Universe,
 Ewha Woman's University, Seoul 120-750, Korea\\
isayev@kipt.kharkov.ua}

\maketitle

\begin{history}
\received{Day Month Year}
\revised{Day Month Year}
\end{history}

\begin{abstract}

Magnetized strange quark stars, composed of strange quark matter
(SQM) and self-bound by strong interactions, can be formed if the
energy per baryon of magnetized SQM is less than that of the most
stable $^{56}$Fe nucleus under the zero external pressure and
temperature. Utilizing the MIT bag model description of magnetized
SQM under charge neutrality and beta equilibrium conditions, the
corresponding absolute stability window in the parameter space of
the theory is determined. It is shown that there exists the maximum
magnetic field strength allowed by the condition of absolute
stability of magnetized SQM. The value of this field,
$H\sim3\cdot10^{18}$~G, represents the upper bound on the magnetic
field strength which can be reached in a strongly magnetized strange
quark star.

\end{abstract}

\ccode{PACS numbers: 25.75.Nq, 21.65.Qr, 98.80.Jk, 95.30.Tg}


\section{Introduction}

It was suggested some time ago that strange quark matter (SQM),
composed of deconfined $u,d$ and $s$ quarks, can be the true ground
state of matter~\cite{AB,W,FJ}. If this conjecture holds true, it
will have important astrophysical implications. In particular,
strange quark stars, composed of SQM and self-bound by strong
interactions, can exist in nature~\cite{I,AFO,HZS}. Also, if SQM is
metastable at zero pressure, it can occur in the high-density core
of a neutron star as a result of the deconfinement phase transition.
In this case, the stability of SQM is provided by the external
pressure from the outer hadronic layers. Then a relevant
astrophysical object is a hybrid star having a quark core and the
crust of hadronic matter \cite{FW}.

Also, an important aspect of the problem is that compact stars can
be strongly magnetized. 
For example, for a special class of neutron stars called magnetars
and, as basically assumed, represented by soft $\gamma$-ray
repeaters (SGRs) and anomalous X-ray pulsars (AXPs), the field
strength can reach values of about $10^{14}$–-$10^{15}$~G
\cite{TD,IShS}. It was also suggested that  magnetized strange quark
stars can be a real source of SGRs or AXPs \cite{CD,OLN}. Strong
magnetic fields of these compact stars can give rise to the
nonlinear QED effects like light bending in the vicinity of a
magnetar \cite{JCAP12JYKim}.  Even stronger magnetic fields up to
$10^{19}$~G may potentially occur in the inner core of a neutron
star \cite{BCP}. As a result, a pulsar can get the large kick
velocity because of the asymmetric neutrino emission in direct Urca
processes in the dense core of a magnetized neutron star
\cite{Ch,ApJ95Vilenkin,HP,PRD99Arras}. The origin of magnetar's
strong magnetic fields is still under discussion, and, among other
possibilities, it is not excluded that this can be due to
spontaneous ordering of hadron \cite{IY,PRC06I,PRC05I,PRC07I}, or
quark \cite{TT} spins in the dense interior of a neutron star.

Thus,  the study of  thermodynamic properties of cold dense matter
in a strong magnetic field is the problem of a considerable interest
\cite{C,BPL,RPPP,IY4,IY10,WSYP}. In particular, the pressure
anisotropy, exhibited in the difference between the longitudinal and
transverse (along and perpendicular to the magnetic field)
pressures, becomes important for strongly magnetized matter
\cite{Kh,IY_PRC11,IY_PLB12,JPG13IY,NPA13SMS}.   In this study, we
consider strongly magnetized SQM within the framework of the MIT bag
model \cite{CJJ}, aiming at determining the parameter space of the
theory for which magnetized SQM is absolutely stable, i.e., its
energy per baryon is less than that of the most stable nucleus
$^{56}$Fe under the zero external pressure and temperature. For the
parameters from this absolute stability window, the formation of a
strongly magnetized strange quark star is possible. Otherwise, a
strongly magnetized hybrid star can be formed. In the latter case,
in strong magnetic fields $H>H_{th}$ (with
$10^{17}<H_{th}\lesssim10^{18}$~G) the transverse pressure increases
with the magnetic field while the longitudinal pressure decreases
\cite{JPG13IY}.  The upper bound on the magnetic field in the quark
core of a hybrid star is determined by the critical field $H_c$
beyond which the longitudinal pressure becomes negative resulting in
the appearance of the longitudinal instability. There exists the
upper bound on the magnetic field strength in a strongly magnetized
strange quark star as well. It is related to the determination of
the absolute stability window for strongly magnetized SQM and
finding the maximum magnetic field strength allowed by the condition
of absolute stability.  Also, in this research we study how the
absolute stability window and upper bound on the magnetic field
strength are affected by varying the strange quark current mass
$m_s$, taking into account scattering in the $m_s$ values existing
in the literature.

\section{Numerical results and discussion}

For the details of the formalism one can address to Ref.
\citen{JPG13IY}. We use the simplified variant of the MIT bag model,
in which quarks are considered as free fermions moving inside a
finite region of space called a "bag". The effects of the
confinement are implemented by introducing the bag pressure $B$,
providing an extra constant energy per unit volume inside the bag.

First, we will determine the absolute stability window of magnetized
SQM, subject to charge neutrality and beta equilibrium conditions,
at zero temperature. In the MIT bag model, the equilibrium
conditions for magnetized
SQM 
in terms of the longitudinal $p_l$ and transverse $p_t$ pressures
read
\begin{align}
p_{l}&=-\Omega_H=0,\label{pl}\\
p_{t}&=-\Omega_H+H\frac{\partial\Omega_H}{\partial H}=0, \label{pt}
\end{align} where $$\Omega_H=\sum_{i=u,d,s,e}\Omega_i+\frac{H^2}{8\pi}+B$$ is the total thermodynamic
potential of the system  including the magnetic field contribution.
With account of Eq.~\p{pl} and at nonzero magnetic field, Eq.~\p{pt}
is reduced to $\frac{\partial\Omega_H}{\partial H}=-\frac{\partial
p_l}{\partial H}=0$. The last equation explicitly reads
\begin{equation}\label{part_pt}
    M-\frac{H}{4\pi}-\frac{\partial B}{\partial H}=0,
\end{equation}
where $M=-\sum_i (\frac{\partial\Omega_i}{\partial H})_{\mu_i}$ is
the matter magnetization and we  assume that the bag pressure can
depend on the magnetic field. For the field-independent bag
pressure, Eq.~\p{part_pt} would be difficult 
to satisfy, because it would mean that the response of the system to
an external magnetic field would considerably exceed the
paramagnetic response, that is
hard to expect without spontaneous spin ordering in the system. 

In order to be absolutely stable, the energy per baryon of
magnetized SQM should be less than the energy per baryon  of the
most stable $^{56}$Fe nucleus under the equilibrium
conditions~\p{pl}, \p{pt}:
\begin{equation}\frac{E_m}{\varrho_B}\leqslant\epsilon_H(^{56}\mathrm{Fe}),
\label{genst}\end{equation} where $E_m\equiv E-\frac{H^2}{8\pi}$ is
the matter part of the total energy density $E$ of the system,
$\varrho_B$ is the total baryon number density. From the equilibrium
condition \p{pl}, one gets
\begin{equation}
B=-\sum_{i=u,d,s,e}\Omega_i-\frac{H^2}{8\pi}. \label{bag}
\end{equation}

The total energy density $E$ of the system, with account of
Eq.~\p{bag}, is given by \cite{JPG13IY}
\begin{equation}
E=\sum_{i=u,d,s,e}
    \mu_i\varrho_i,
\end{equation}
where $\varrho_i=-(\frac{\partial\Omega_i}{\partial\mu_i})_H$ is the
number density for fermions of $i$th species with the chemical
potential $\mu_i$. The stability constraint \p{genst} then reads
 \begin{equation}\label{stc}
    \biggl.\frac{E_m}{\varrho_B}\biggr|_{SQM}=\frac{1}{\varrho_B}{\sum_{i=u,d,s,e}
    \mu_i\varrho_i}-\frac{H^2}{8\pi\varrho_B}\leqslant\epsilon_H(^{56}\mathrm{Fe}).
 \end{equation}

For the rough estimate, one can regard $^{56}$Fe nucleus as a system
of noninteracting nucleons, and, therefore, magnetic fields
$H>10^{20}$~G are necessary in order to significantly alter its
energy per nucleon \cite{BPL}. Since we will consider magnetic
fields $H<5\cdot10^{18}$~G, we use the approximation
$\epsilon_H(^{56}\mathrm{Fe})\approx
    \epsilon_0(^{56}\mathrm{Fe})=930\,\mathrm{MeV}$.

     Thus, in order to find the upper bound $B_u$ on the bag pressure $B$
for magnetized SQM to be absolutely stable, it is necessary, first,
to find the fermion species chemical potentials $\mu_i$
($i=u,d,s,e$) from the  constraint~\p{stc}, taken with the equality
sign, and charge neutrality and chemical equilibrium conditions:
\begin{align}2\varrho_u-\varrho_d-\varrho_s-3\varrho_{e^-}=0,\label{cnc}\\
\mu_d=\mu_u+\mu_{e^-},\label{mud}\\
\mu_d=\mu_s. \label{ds}
\end{align}
Then the upper bound $B_u$ on the bag pressure from the absolute
stability window can be found from Eq.~\p{bag}. Note that, as a
solution of Eq.~\p{bag}, the function $B_u(H)$ will identically
satisfy to Eq.~\p{pt}.

At $H=0$, in turn, the lower bound $B_l$ on the bag pressure from
the absolute stability window can be established from the
experimental observation that two-flavor quark matter, consisting of
$u$ and $d$ quarks, is less stable compared to $^{56}$Fe nucleus at
zero external pressure and temperature \cite{FJ}. Obviously, that
requirement is preserved at not too strong magnetic fields, at
least, up to $H\sim10^{17}$~G, when the impact of a magnetic field
on quark matter  properties is insignificant \cite{C,JPG13IY}:

\begin{equation}\label{stc1}
    \biggl.\frac{E_m}{\varrho_B}\biggr|_{ud}=\frac{1}{\varrho_B}{\sum_{i=u,d,e}
    \mu_i\varrho_i}-\frac{H^2}{8\pi\varrho_B}\geqslant\epsilon_H(^{56}\mathrm{Fe}).
 \end{equation}

We will also retain the constraint \p{stc1} for stronger magnetic
 fields $H>10^{17}$~G, although, strictly speaking, it is unknown
 from the experimental point of view
 whether two-flavor quark matter is less stable than
$^{56}$Fe nucleus under the equilibrium conditions \p{pl}, \p{pt} in
such strong fields. Nevertheless, as will be shown later, the lower
bound $B_l$, determined in accordance with the constraint~\p{stc1},
  in strong magnetic fields decreases with $H$, and, hence,
becomes less restrictive.

  Before presenting the
absolute stability window in the plane "magnetic field strength --
bag pressure", let us also mention that there is quite a noticeable
gap between the values of the  strange quark current mass $m_s$ used
in many researches on SQM, and that given by PDG. For example, in
Refs. \citen{BCP,RPPP,JPG09Felipe,JPG13IY}, it was used the value
$m_s=150$~MeV while the recent PDG edition  points out the value
$m_s=95\pm5$~MeV\cite{PDG12}. In view of that, we present the upper
bound on $B$ from the absolute stability window for a few values of
$m_s$ in the range from $90$~MeV to $150$~MeV in order to study
quantitatively the effect of varying  $m_s$. For $u$ and $d$ quarks
we use the current masses $m_u=m_d=5$~MeV.

\begin{figure}[tb]
\begin{center}
\includegraphics[height=5.965cm]{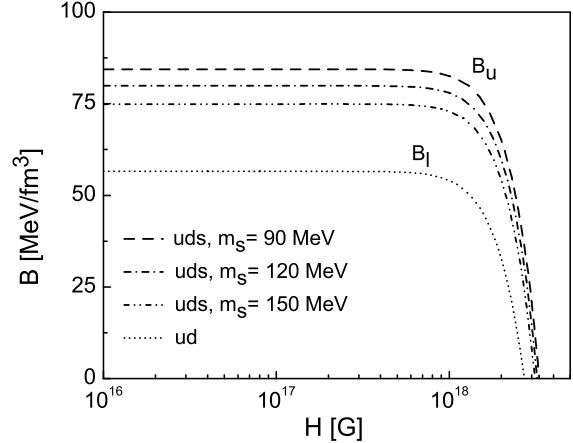}
\end{center}
\vspace{-2ex} \caption{The absolute stability window in the plane
"magnetic field strength - bag pressure" for magnetized SQM at zero
temperature. The upper bound $B_u$ on the bag pressure $B$ is
calculated for varying  $s$ quark current mass $m_s$. }
\label{fig1}\vspace{-0ex}
\end{figure}

Figure~\ref{fig1} shows the upper $B_u$ and lower $B_l$ bounds on
the bag pressure $B$ from the absolute stability window as functions
of the magnetic field strength. It is seen that under increasing the
 $s$ quark current mass $m_s$, the upper bound  $B_u$ decreases. For
example, the maximum value of $B_u$, corresponding to $H=0$ (which
is practically indistinguishable from the value of $B_u$ at
$H=10^{16}$~G) is $B_{u\,max}\approx 84.4$~MeV/fm$^3$ for
$m_s=90$~MeV, $B_{u\,max}\approx 79.9$~MeV/fm$^3$ for $m_s=120$~MeV,
and $B_{u\,max}\approx 74.9$~MeV/fm$^3$ for $m_s=150$~MeV.
 The upper
bound stays, first, practically constant and then, beginning from
the magnetic field strength  $H$  somewhat smaller than $10^{18}$~G,
 decreases with $H$ and becomes, hence, more restrictive. The upper
bound $B_u$ vanishes at $H_{u\,max}\approx3.3\times 10^{18}$~G for
$m_s=90$~MeV, at $H_{u\,max}\approx3.2\times 10^{18}$~G for
$m_s=120$~MeV and at $H_{u\,max}\approx3.1\times 10^{18}$~G for
$m_s=150$~MeV.   In a stronger magnetic field, in order to satisfy
the constraints \p{pl}, \p{pt}, \p{stc}-\p{ds}, the bag pressure had
to become negative, contrary to the constraint $B>0$. This means,
that, under such magnetic fields, the equilibrium conditions \p{pl},
\p{pt}, \p{cnc}-\p{ds} become incompatible with the stability
condition \p{stc} for the positively defined bag pressure. In other
words, under the equilibrium conditions and in magnetic fields
$H>H_{u\,max}$, magnetized SQM becomes less stable than the
$^{56}$Fe nucleus. 

The behavior of the lower bound $B_l$ with the magnetic field is
similar to that of the upper bound $B_u$. At $H=0$, the lower bound
has its maximum value $B_{l\,max}\approx56.5$~MeV/fm$^3$ (which
almost coincides with the value of $B_l$ at $H=10^{16}$~G). It stays
practically constant till magnetic fields somewhat smaller than
$10^{18}$~G, beyond which $B_l$ decreases with $H$ and becomes,
hence, less restrictive. The lower bound $B_l$ vanishes at
$H_{l\,0}\approx 2.7\times10^{18}$~G. Under the equilibrium
conditions and in the fields $H>H_{l\,0}$, the lower bound $B_l$
would be negative. Because of the positiveness of the bag pressure,
the inequality $B>B_l$ would be fulfilled always in the fields
$H>H_{l\,0}$.   Thus, in order
 magnetized SQM would be absolutely stable, the magnetic field strength
 should satisfy the constraint $H<H_{u\,max}$. In fact, the value $H_{u\,max}$
represents the upper bound on the magnetic field strength which can
be reached in a magnetized strange quark star.

Note that the absolute  stability window of magnetized SQM  in the
MIT bag model was  studied earlier in Ref. \citen{JPG09Felipe}.
However, in that study the pure magnetic field contribution
$\frac{H^2}{8\pi}$ (the Maxwell term) to  the longitudinal $p_l$ and
transverse $p_t$ pressures in Eqs. \p{pl}, \p{pt} was missed.
Because the Maxwell term enters with different sign to the pressures
 $p_l$ and $p_t$, it cannot be excluded by the
redefinition of the bag pressure. This term becomes important just
in the range of strong magnetic fields $H>H_{th}$ with
$10^{17}<H_{th}\lesssim10^{18}$~G \cite{JPG13IY}. By this reason, in
Ref. \citen{JPG09Felipe} the bag pressure from the absolute
stability window was obtained as only weakly field dependent, and no
any constraint on the magnetic field strength was set  from the
requirement of absolute stability of magnetized SQM.

\begin{figure}[tb]
\begin{center}
\includegraphics[height=5.965cm]{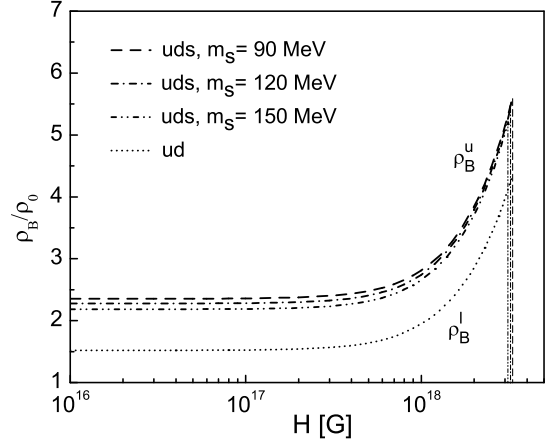}
\end{center}
\vspace{-0ex} \caption{The baryon number density $\varrho_B$ vs.
magnetic field strength $H$ for magnetized SQM (three upper curves)
and magnetized two-flavor quark matter, determined under the
respective equilibrium conditions for $E_m/\varrho_b=930$~MeV. The
bounding vertical lines on the right correspond to $H=H_{u\,max}$.}
\label{fig2n}\vspace{-0ex}
\end{figure}

It is worthy to note at this point that in the case of magnetized
hybrid stars the stability of the quark core is provided by the
gravitational pressure from the outer hadronic layers, and the bag
pressure can be field-independent in the metastable state. Then,
without evidently counting the gravitational pressure in the
equilibrium conditions, the longitudinal $p_l$ and transverse $p_t$
pressures in the quark core would be positive. Under strong magnetic
fields, the longitudinal pressure $p_l$ decreases with the magnetic
field while the transverse pressure $p_t$ increases. There exists
the critical magnetic field beyond which the longitudinal pressure
$p_l$ becomes negative resulting in the appearance of the
longitudinal instability in magnetized quark matter \cite{JPG13IY}.
In the astrophysical context, this would mean the occurrence of the
gravitational collapse of a hybrid star along the magnetic field, if
the magnetic field strength exceeds the critical one. The magnitude
of the critical field for the appearance of the longitudinal
instability represents the upper bound on the magnetic field
strength in the interior of a hybrid star. Thus, in both instances
of a strongly magnetized strange quark star and a strongly
magnetized hybrid star there exists the upper bound on the magnetic
field strength, related to the issue of the star's stability, but
the concrete mechanism responsible for the appearance of the
instability   is different in each case.

Figure~\ref{fig2n} shows the total baryon number density, normalized
on the nuclear saturation density $\varrho_0=0.16$~fm$^{-3}$, as a
function of the magnetic field strength for magnetized SQM and
magnetized two-flavor quark matter, calculated under the respective
equilibrium conditions for $E_m/\varrho_b=930$~MeV.  In fact, the
corresponding lines $\varrho_B^u(H)$ and $\varrho_B^{\,l}(H)$
represent the upper and lower bounds on the total
baryon number density 
to ensure the absolute stability of magnetized SQM, assuming that,
under the equilibrium conditions, two-flavor quark matter remains to
be less stable than $^{56}$Fe nucleus in strong magnetic fields.
 On the right, the
absolute stability window is bound by the straight line
$H=H_{u\,max}$.  The corresponding maximum values
$\varrho_{B\,max}^u$ and $\varrho_{B\,max}^l$ of the upper and lower
bounds on the total baryon density are
$\varrho_{B\,max}^u\approx5.6\varrho_0$ and
$\varrho_{B\,max}^l\approx4.4\varrho_0$ at $m_s=90$~MeV,
$\varrho_{B\,max}^u\approx5.4\varrho_0$ and
$\varrho_{B\,max}^l\approx4.3\varrho_0$ at $m_s=120$~MeV,
$\varrho_{B\,max}^u\approx5.2\varrho_0$ and
$\varrho_{B\,max}^l\approx4.1\varrho_0$ at $m_s=150$~MeV.
 It is seen that the upper $\varrho_B^u$ and
lower $\varrho_B^l$ bounds for the allowable total baryon number
density $\varrho_B$ from the absolute stability window stay
practically constant till the magnetic field strength somewhat
smaller than $10^{18}$~G, and then increase till the corresponding
maximum value.  Also, the increase of the current mass $m_s$ leads
to the decrease of the upper bound on $\varrho_B$, e.g., at $H=0$
 (giving nearly the same results as at $H=10^{16}$~G),
$\varrho_{B\,min}^u\approx2.4\varrho_0$ for $m_s=90$~MeV,
$\varrho_{B\,min}^u\approx2.3\varrho_0$ for $m_s=120$~MeV,
$\varrho_{B\,min}^u\approx2.2\varrho_0$ for $m_s=150$~MeV. The lower
bound on $\varrho_B$ at $H=0$ is
$\varrho_{B\,min}^l\approx1.5\varrho_0$. Note that magnetic fields
$H\gtrsim10^{18}$~G strongly affect the upper and lower bounds on
the baryon number density from the absolute stability window, unlike
to the results of Ref. \citen{JPG09Felipe}, where this impact was
found to be modest.

In conclusion, we have considered magnetized strange quark stars,
composed of SQM, within the framework of the MIT bag model aiming at
determining the absolute stability window in the parameter space of
the theory for which magnetized strange quark stars can be formed.
To that end, we have determined the domain of magnetic field
strengths, bag pressures, and total baryon number densities, for
which the energy per baryon of magnetized SQM is less than that of
the most stable $^{56}$Fe nucleus under the zero external pressure
conditions \p{pl}, \p{pt} and vanishing temperature.
  In fact, this
requirement sets the upper bound on the parameters from the absolute
stability window. The lower bound on the parameters from the
absolute stability window is determined from the constraint that
magnetized two-flavor quark matter under equilibrium conditions
\p{pl}, \p{pt} and zero temperature should be less stable than the
most stable $^{56}$Fe nucleus. This constraint is extended from weak
terrestrial magnetic fields, where it has direct experimental
confirmation, to possible strong magnetar interior magnetic fields
$H>10^{17}$~G, where such confirmation is wanting.  An important
feature of our consideration is that, unlike to some of the previous
studies, in the zero pressure conditions \p{pl}, \p{pt} the pure
magnetic field contribution (the Maxwell term) to the transverse and
longitudinal pressures has been taken into account. It has been
shown that there exists the magnetic field strength at which the
upper bound on the bag pressure from the absolute stability window
vanishes. In fact, the value of this field, $H_{u\,
max}\sim3\cdot10^{18}$~G, represents the upper bound on the magnetic
field strength, which can be reached in a strongly magnetized
strange quark star. Also, we have studied  how the absolute
stability window and upper bound on the magnetic field strength are
affected by varying the strange quark current mass $m_s$. It has
been clarified that the increase of the current mass $m_s$ leads to
the decrease of the upper bound on the bag pressure and total baryon
number density from the absolute stability window, and to the
decrease of the upper bound on the magnetic field strength in
magnetized strange quark stars as well.

Note that obtained here an estimate of the upper bound on the
magnetic field strength in magnetized strange quark stars may be
further improved by utilizing a more elaborated version of the MIT
bag model, e.g., with taking into account the effects of short-range
quark interactions or color superconductivity
\cite{arxiv10Lattimer,PRD13Wen}. Nevertheless, the given analysis
provides, definitely, the correct order of magnitude of the upper
bound on the magnetic field strength in strange quark stars, and is
acceptable for getting the rough estimate of this quantity.

\section*{Acknowledgment}

The author would like to thank Ewha Woman's University for
hospitality and support during his stay at Seoul.



\end{document}